\documentclass[11pt]{article}

\usepackage[margin=1.2in]{geometry} 

\usepackage[utf8]{inputenc}
\usepackage{amsmath, amssymb, graphicx}

\usepackage{url}
\usepackage{booktabs}
\usepackage{xurl}
\usepackage{hyperref}

\usepackage[super,sort&compress]{natbib} 
\title{Symptom clusters in Long COVID in the UK: prospective community-based cohort study using unsupervised machine learning}

\author{
  \parbox{\textwidth}{\centering
   Jasmine Aherne$^{1*}$, In\^{e}s Henriques-Cadby$^{1,2}$, and Thomas House$^{1}$
  } \\[3ex]
  \parbox{0.9\textwidth}{\centering\small
    $^1$Department of Mathematics, University of Manchester, Manchester, United Kingdom \\
    $^2$School of Medicine and Population Health, The University of Sheffield, Sheffield, UK \\[2ex]
    $^*$Corresponding author email: jasmine.aherne@manchester.ac.uk\\[3ex]
  }
}
\date{\today} 

\begin{document}
\maketitle

\begin{abstract}
\textbf{Background:} Long COVID is a condition usually defined by persisting symptoms following infection by the SARS-CoV-2 virus beyond the acute phase of infection. The condition has a significant impact on healthcare systems, the economy, and most importantly the individuals living with it. Due to the diverse and extensive symptomatology of long COVID, symptom co-occurrence tracking can be used to capture patient experiences and improve understanding, diagnosis, and management.  Here, we leverage the UK Office for National Statistics (ONS) COVID-19 Infection Survey (CIS), which contains information on long COVID but has not previously been used to investigate symptom co-occurrence.
\textbf{Methods:} The CIS was run between April 2020 and March 2023. It has a longitudinal, prospective community cohort design and a study population comprising a nationally representative sample of households with over 500,000 participants. On February 3, 2021, the ONS launched a new CIS question evaluating self-reported symptom persistence of 23 long COVID symptoms post self-reported COVID-19 infection. We use Jaccard symptom-by-symptom distance matrices, derived from the binary survey responses of  the presence of each symptom. Three methods are used to visualise symptom co-occurrence from the distance matrices: heatmaps, non-metric multidimensional scaling, and agglomerative hierarchical clustering with complete linkage. We split our analysis into two parts, first looking at all long COVID survey responses (\(n\) = 207,319) and then looking at responses stratified by time-since-onset of long COVID up to 24 months (\(n\) = 30,224 at zero months-since-onset).
\textbf{Results:} We find higher symptom co-occurrence in prolonged long COVID. We also find clusters of neurological/systemic symptoms, gastrointestinal symptoms, and respiratory symptoms, with separability in symptom co-occurrence by organ system becoming less pronounced as time-since-onset of long COVID increases. Shortness of breath, weakness/tiredness, and muscle ache present as core symptoms of long COVID.
\textbf{Conclusions:} We demonstrate that the symptom experience of long COVID evolves from early stages to late stages with higher symptom burden and multi-systemic presentation. This indicates that the number of symptoms could define severity of later long COVID. We also find three possible phenotypes of early long COVID, from neurological/systemic, respiratory, and gastrointestinal symptom groups, which are consistent with the broader literature.
\end{abstract}

\section{Background}
Long COVID is typically considered to refer to a condition in which symptoms following infection by the SARS-CoV-2 virus persist beyond the acute phase of infection. These sequelae can be multi-systemic and highly heterogeneous among patients \cite{NICE_LC}. In 2021, the World Health Organization stated that symptoms must last more than 3 months after acute COVID-19 and can fluctuate in severity \cite{WHO2021}; however, a uniformly agreed definition of long COVID still does not exist \cite{Gomez-Bravo2025}. 

Six years on from the first reported case of COVID-19 in December 2019, data from the 2026 GP Patient Survey, which surveyed a nationally representative sample of patients aged 16 and over registered with a GP practice in England, revealed that 3.5\% of people describe themselves as currently living with long COVID (down from 4.2\% in 2025) and 9.1\% say they were unsure (down from 9.5\% in 2025) \cite{NHSEngland2026}. Furthermore, as evidenced by a weekly peak of 3,507 cases of hospital-confirmed COVID-19 in England during the second half of 2025, COVID-19 continues to circulate and impact healthcare systems \cite{UKHSA_COVIDreport}. Finally, \citet{Kwon2025} estimated that the annual cost of long COVID up to March 7th, 2024 was £8.1 billion in England and Scotland, similar to the annual cost of stroke in England. On an international scale, the Organisation for Economic Co-operation and Development (OECD) estimates that the direct annual healthcare cost of long COVID will be approximately £11 billion over the next decade, with indirect costs totalling £101 billion per year, across all member nations of the OECD and the European Union \cite{OECD2026}. Given the sustained significant prevalence of long COVID, the continued presence of SARS-CoV-2, the projected continued cost of long COVID, and the possibility of future viral outbreaks, ongoing research into long COVID remains a public health priority. This is underscored by other post-acute infection syndromes which present with similar symptoms to long COVID and historically account for significant healthcare utilisation \cite{Choutka2022}.

Due to the diverse and extensive symptomatology of long COVID, which demonstrates heterogeneity in severity and type \cite{NICE_LC}, symptom co-occurrence tracking is essential to capture patient experiences, improve understanding and diagnosis, and facilitate targeted management. This tracking is particularly critical given that individuals with long COVID frequently experience functional limitations, restrictions to social participation, and economic burdens \cite{Wang2024} alongside worse health-related quality of life scores than those with other medical conditions including heart failure, multiple sclerosis, and end-stage renal failure \cite{Carlile2024}.

A growing body of literature exists on symptom clustering in long COVID. The most recent systematic review found that 25\% of studies clustered by organ system in the symptom space, while 46.9\% explored clustering in the patient space to find potential patient phenotypes \cite{Wang2026}. In the 16 studies that clustered by organ system, the most common symptom clusters were neurological, respiratory, and gastrointestinal. On the other hand, those looking at potential patient phenotypes found olfactory–gustatory dysfunction (loss of smell and taste), fatigue–dyspnea (weakness/tiredness with shortness of breath), and fatigue–cognitive/neurological symptom clusters. Another systematic review looking at the prevalence of symptoms corroborates these most frequent organ system symptom clusters while also demonstrating that symptom presentation changes as time-since-onset of long COVID increases \cite{Hu2025}. In particular, \citet{Hu2025} found that as time-since-onset increases, fatigue becomes a core symptom, while neurological, circulatory, and gastrointestinal symptoms drive phenotypes. We note that only 41\% and 32\% of the included studies in the respective systematic reviews are prospective cohort studies, the strongest observational study design \cite{Browner2023}. This means the majority of the included studies relied on retrospective or cross-sectional frameworks and were vulnerable to increased biases.

Considerable research has also been conducted looking at symptom co-occurrence in acute COVID-19, for example by \citet{Sudre2021}, \citet{Millar2022}, and, in the COVID-19 Infection Survey (CIS), \citet{Fyles2023}. Whilst the CIS is a major data source, with its unprecedented sample size for a prospective community-based cohort design being core a strength, we believe it currently remains under-utilised for the analysis of long COVID. In this paper, we will leverage the survey data to explore long COVID symptom clusters.

\section{Methods}

\subsection{Study data and design}
This work uses the COVID-19 Infection Survey (study number: ISRCTN21086382) which was run by the Office for National Statistics (ONS) in the United Kingdom (UK) \cite{CIS_dataset, Oxford_protocols}. The CIS utilised a longitudinal, prospective community cohort design from April 2020 to March 2023. The study population comprised a nationally representative cohort of households and had over 500,000 participants. Data collection consisted of swab tests (for individuals aged \(\geq 2\) years old), a few short questions, and, a sample of blood (from those aged 16 years or older in a further randomly selected subset of households). Follow ups were in general every week for 4 weeks and then monthly. Participants were compensated for taking part.

The ONS first started reporting data on long COVID on 16 December 2020 \cite{ONS_LC_initial}, initially looking at participants whose symptoms of COVID-19 had extended beyond 5 and 12 weeks. On February 3, 2021 the ONS officially launched a new CIS question on long COVID  \cite{ONS_LC_April21}. The question evaluated self-reported symptom persistence exceeding four weeks post self-reported COVID-19 infection, whether the symptoms were affecting their ability to carry out daily activities, and which of the 23 symptoms listed in Table \ref{Table1} they were experiencing that were not attributable to anything else.

\begin{table}[htb]
\caption{Long COVID symptoms included in the Office for National Statistics COVID-19 Infection Survey.}
\label{Table1}
\begin{tabular}{p{0.2\linewidth} p{0.7\linewidth}}
\toprule
\textbf{Symptom type} & \textbf{Symptom} \\
\midrule
Neurological &
Difficulty concentrating, memory loss/confusion, worry/anxiety,
low mood/loss of interest, loss of smell, loss of taste,
trouble sleeping, headache, vertigo/dizziness \\
Respiratory &
Cough, shortness of breath, noisy breathing, sore throat,
runny nose/sneezing \\
Systemic &
Fever, weakness/tiredness, muscle ache, palpitations, chest pain \\
Gastrointestinal &
Diarrhoea, nausea/vomiting, loss of appetite, abdominal pain \\
\bottomrule
\end{tabular}

\vspace{1.5ex}
\noindent{\footnotesize{Symptoms were grouped into organ systems for descriptive purposes in this work.}}
\end{table}

\subsection{Data preparation}
This analysis includes every survey response where participants responded positively to the self-reported long COVID question. This means that there are some included participants who had not received a positive test for COVID-19, however, the percentage of these participants was highest in those who suspected they had experienced COVID-19 prior to the introduction of national testing \cite{ONS_LC_May22}. Hence, we include all participants to minimise selection bias.

\begin{figure}[htb]
\centering
\includegraphics[width=0.7\textwidth]{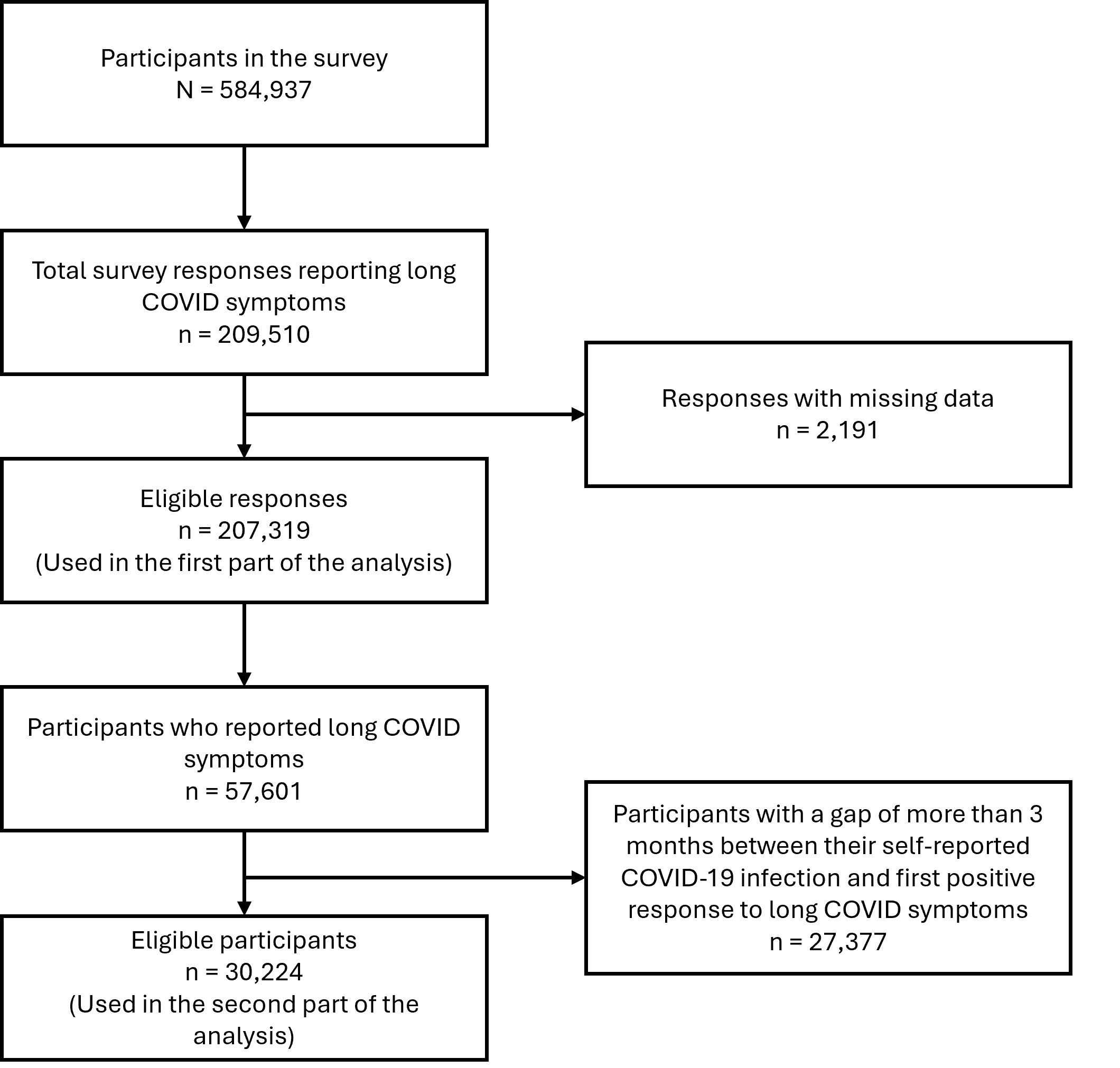}
\caption{Inclusion and exclusion criteria for the COVID-19 Infection Survey responses (r) and participants (n).}
\label{Figure1}
\end{figure}

For the first part of this analysis we included every positive long COVID response, meaning that in some cases there were multiple responses per participant. In the second part of the analysis, we stratified responses by time-since-onset of long COVID. First, we used each participants' first positive response to long COVID. We limited this analysis to participants who indicated that they suspected they had had COVID-19 within 3 months of their first positive response to long COVID. This prevents the inclusion of participants who had developed long COVID long before the introduction of the long COVID question to the survey and participants who had disclosed their long COVID long after it had developed. Inclusion of these participants would include first positive responses to long COVID that are not actually near the start of the participants' experience of long COVID (these participants are included in the first part of the analysis). We did not restrict further than this, leaving a large number of eligible participants with a straightforward link between acute COVID-19 and long COVID. We then stratified these participants' responses by time since their first long COVID response (at three months, six months, 12 months, 18 months, and 24 months). A full workflow is given in Figure \ref{Figure1}. As shown in Table \ref{Table2}, the number of participants decreases across successive stratified time points. This could be for a number of factors including survey participants recovering, survey participant attrition (although this was reported up to the end of July 2022 to be low \cite{ONS_Attrition}), or survey participants developing long COVID closer to the later stages of the survey. Table \ref{Table2} also presents participants characteristics.

\begin{table*}[t]
\caption{Characteristics of the COVID-19 Infection Survey sample.}
\label{Table2}
\begin{tabular*}{\textwidth}{@{\extracolsep{\fill}}lc*{6}{c}@{}}
\toprule
&
&
\multicolumn{6}{c}{\textbf{Month-since-onset}} \\
\cmidrule{3-8}
\textbf{Variable} &
\textbf{All} &
\textbf{0} &
\textbf{3} &
\textbf{6} &
\textbf{12} &
\textbf{18} &
\textbf{24} \\
\midrule
\multicolumn{8}{@{}l}{\textit{Sample size}} \\
$n$ & 207319 & 30224 & 5717 & 3787 & 1606 & 644 & 45 \\
\addlinespace
\multicolumn{8}{@{}l}{\textit{New cases by year}} \\
2021 (Feb--Dec) & -- & 9329 & 1389 & -- & -- & -- & -- \\
2022 & -- & 19407 & 3832 & -- & -- & -- & -- \\
2023 (Jan--Mar) & -- & 1488 & 496 & -- & -- & -- & -- \\
\addlinespace
\multicolumn{8}{@{}l}{\textit{Sex breakdown}} \\
Female & 126146 & 11547 & 2151 & 1411 & 585 & 247 & 16 \\
Male & 81173 & 18677 & 3566 & 2376 & 1021 & 397 & 29 \\
\addlinespace
\multicolumn{8}{@{}l}{\textit{Age breakdown}} \\
2 -- 11 & 1888 & 526 & 44 & 24 & * & 0 & 0 \\
12 -- 16 & 5483 & 1055 & 190 & 115 & * & * & 0 \\
17 -- 24 & 6396 & 1141 & 152 & 88 & 52 & * & * \\
25 -- 34 & 13397 & 2398 & 397 & 232 & 86 & 32 & * \\
35 -- 49 & 51400 & 7788 & 1482 & 1050 & 451 & 160 & * \\
50 -- 69 & 98050 & 12671 & 2648 & 1790 & 788 & 364 & 29 \\
70 + & 30705 & 4645 & 804 & 488 & 178 & 73 & * \\
\addlinespace
\multicolumn{8}{@{}l}{\textit{Prior long term health condition(s)}} \\
No & 121198 & 20388 & 3441 & 2206 & 920 & 343 & 20 \\
Yes & 86121 & 9836 & 2276 & 1581 & 686 & 301 & 25 \\
\addlinespace
\multicolumn{8}{@{}l}{\textit{Number of COVID-19 vaccinations received}} \\
0 & 173913 & 25713 & 4646 & 2926 & 1112 & 446 & 32 \\
1 & 8890 & 1158 & 327 & 267 & 227 & 177 & 13 \\
2 & 12014 & 967 & 222 & 189 & 151 & 21 & 0 \\
3 or more & 12502 & 2386 & 522 & 405 & 116 & 0 & 0 \\
\addlinespace
\multicolumn{8}{@{}l}{\textit{Mean number of symptoms experienced}} \\
All participants & 5.04 & 4.13 & 5.03 & 5.55 & 5.83 & 7.22 & 8.18 \\
\bottomrule
\end{tabular*}

\vspace{1.5ex}
\noindent{\footnotesize{The first column summarises the total eligible survey responses; the remaining columns give the eligible responses stratified by month-since-onset of long COVID. Counts suppressed due to statistical disclosure control rules are indicated by *.}}
\end{table*}

The included symptoms were inputted by response in the data with a 0 indicating the absence of a symptom and a 1 indicating the presence of a symptom. Missing data was minimal: the majority of cases of missingness being where a survey participant had given a positive response to long COVID but had not given a response to any of the symptoms. For this reason responses with missing data were excluded.

\subsection{Statistical methods}

\subsubsection{Jaccard distance}
For any two symptom vectors with binary elements corresponding to the presence (1) or absence (0) of the symptoms respectively at each case, the Jaccard distance is given by
\begin{equation}
    D_J = 1 - \frac{M_{11}}{M_{11} + M_{10} + M_{01}}
    \label{Equation1}
\end{equation}
where \(M_{11}\) is the number of cases in which both symptoms are experienced, \(M_{10}\) is the number of cases in which the first symptom is experienced whilst the second is not, and \(M_{01}\) is the number of cases where the first symptom is not experienced whilst the second is. \(M_{00}\) would be the number of cases in which both symptoms are not experienced, but is excluded here and is not considered informative: two symptoms being absent in a patient does not make those symptoms similar in a clinical sense and is not predictive of symptom co-occurrence. Hence, the Jaccard distance between two symptoms can be thought of as the proportion of cases not experiencing both symptoms given the number of cases experiencing at least one of the symptoms. We see that if \(D_J = 0\) the symptoms are always experienced together whilst if \(D_J = 1\) the symptoms are never experienced together. Obtaining \(D_J\) for all pairs of symptoms allows us to construct the Jaccard distance matrix of symptoms, where the \((i,j)\)-th element corresponds to the Jaccard distance between a given symptom \(i\) and symptom \(j\).

\subsubsection{Non-metric multidimensional scaling}
Non-metric multidimensional scaling (NMDS) is an ordination technique used to represent high dimensional data in a low dimensional space \cite{Kruskal1964theory,Kruskal1964method}. Starting with a distance matrix of symptoms, the algorithm arranges the symptoms such that those which are close in the high dimensional space are close in the low dimensional space by preserving the rank order of the original distances between symptoms. We let \(X\) be the data matrix in the high dimensional space that has \(n\) rows (corresponding to the number of participants or responses in each analysis) and \(p\) columns (corresponding to the number of symptoms). The first run of the algorithm typically allocates the lower dimensional symptom co-ordinates randomly. The goodness of fit of this low dimensional arrangement is given by the stress
\begin{equation*}
    S = \sqrt{\frac{\sum_{i=1}^{p-1} \sum_{j=i+1}^{p} (d_{ij} - \tilde{d}_{ij})^2}{\sum_{i=1}^{p-1} \sum_{j=i+1}^{p} d_{ij}^2}}\,,
\end{equation*}
where
\begin{equation*}
    d_{ij} = \sqrt{\sum_{k=1}^{n} (x_{ki} - x_{kj})^2}
\end{equation*}
is the euclidean distance between each symptom in the low co-ordinate space and the \(\tilde{d}_{ij}\) are chosen by monotonic regression (they equal \(d_{ij}\) where the rank order between the \(d_{ij}\) preserves the rank order of the distances between points in the high dimensional space and are an average of the \(d_{ij}\) where the rank order is violated). Simply, stress is a measure of how much the distance between symptoms in the low dimensional space deviates from ideal values that preserve the original rank order of the symptoms. It shows how well the co-ordinates of the symptoms in the low dimensional space match the original symptom distance matrix. 

Since the formula for stress is a function of the coordinates of the symptoms in the low dimensional space, its partial derivatives can be used to calculate a gradient and be used in the method of steepest descent to minimise the stress score. The coordinates of the symptoms are adjusted in the direction of the negative gradient. The stress is then recalculated with the updated coordinates of the symptoms, and the process continues iteratively until a convergence criterion is met, which is usually when the reduction in stress between iterations falls below a predefined threshold. To ensure a global minimum is found, multiple runs of the algorithm are completed with the one giving the lowest stress selected as the final model. Typically, stress below \(0.05\) is deemed good, \(0.1-0.2\) is deemed fair, and above \(0.2\) is deemed poor \cite{Schumacker2016}.

As the method is non-metric, the distance between the symptoms in the low dimensional space is meaningless except to say that a symptom \(i\) is closer to a symptom \(j\) than a symptom \(k\), but it does mean that the method is robust to complex, noisy, and non-linear data. We can use a plot of the resulting co-ordinates of symptoms in the low dimensional space to look for frequently co-occurring symptoms, where symptoms that co-occur more frequently will be located closer together, and in the exploration of clusters. 

We performed NMDS using the \texttt{metaMDS} function in the R package \texttt{vegan} \cite{RVegan}.

\subsubsection{Agglomerative hierarchical clustering}
Agglomerative hierarchical clustering is a bottom-up approach to clustering \cite{James2021}. Starting with a distance matrix of symptoms, the algorithm initially assigns each symptom to its own cluster and then merges the pair of clusters that are closest. It then assigns the distance between the new cluster and the other clusters according to a chosen linkage method. This is continued iteratively until all symptoms are grouped into a single cluster. 

We chose to use complete linkage, the maximum distance between a symptom in the new cluster to each other cluster. This means that to merge two groups of symptoms the most distant symptoms must be close enough, preventing clusters from being too spread out. These distances at which each cluster of symptoms are merged can be visualised in a dendrogram plot. 

The dendrogram is a tree-like plot that details at which distances clusters are merged. We can use this plot to look for frequently co-occurring symptoms as merges at smaller distances indicate symptoms that are more frequently experienced together, whilst merges at bigger distances indicate symptoms less frequently experienced together. Similarly, cutting the dendrogram at a pre-defined height allows for the exploration of clusters.

We choose both NMDS and agglomerative hierarchical clustering as the spatial visualisation and cluster assignment results from each method respectively complement each other. Furthermore, they allow for comparison and ensure findings are not artefacts of one method.

\section{Results}

\subsection{Jaccard distance}
\begin{figure*}[htb]
    \centering
    \includegraphics[width=0.9\textwidth]{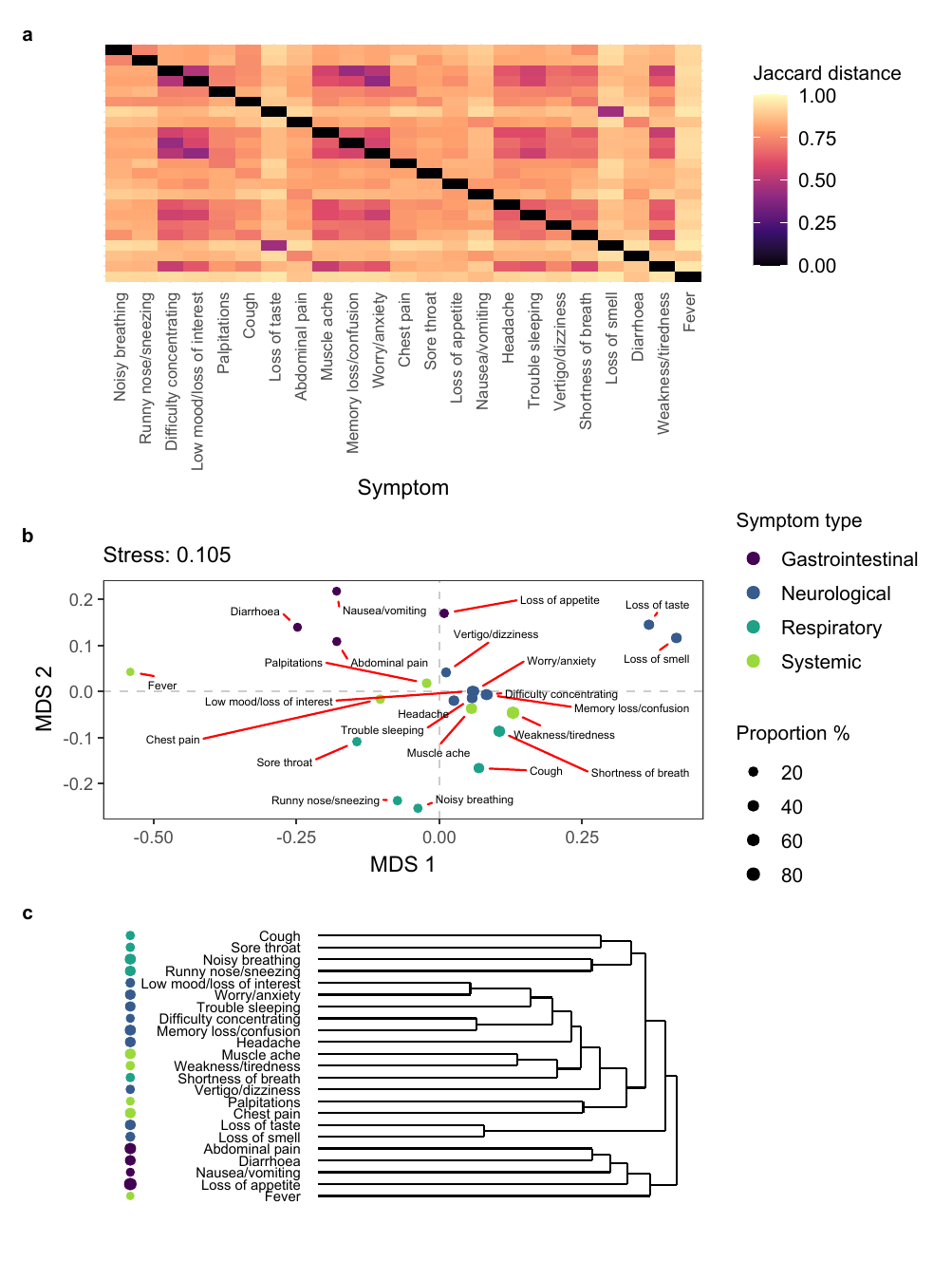}
    \caption{Results from the first part of the analysis including all responses reporting long COVID symptoms. Panel (a) shows the Jaccard distance matrix visualised as a heatmap, panel (b) shows the non-metric multidimensional scaling results, and panel (c) shows the dendrograms visualising the agglomerative hierarchical clustering results using complete linkage. Points in (b) and (c) are colour-coded by symptom type and sized to represent the proportion of participants who experienced said symptom.}
    \label{Figure2}
\end{figure*}

\begin{figure*}[htb]
    \centering
    \includegraphics[width=0.9\textwidth, keepaspectratio]{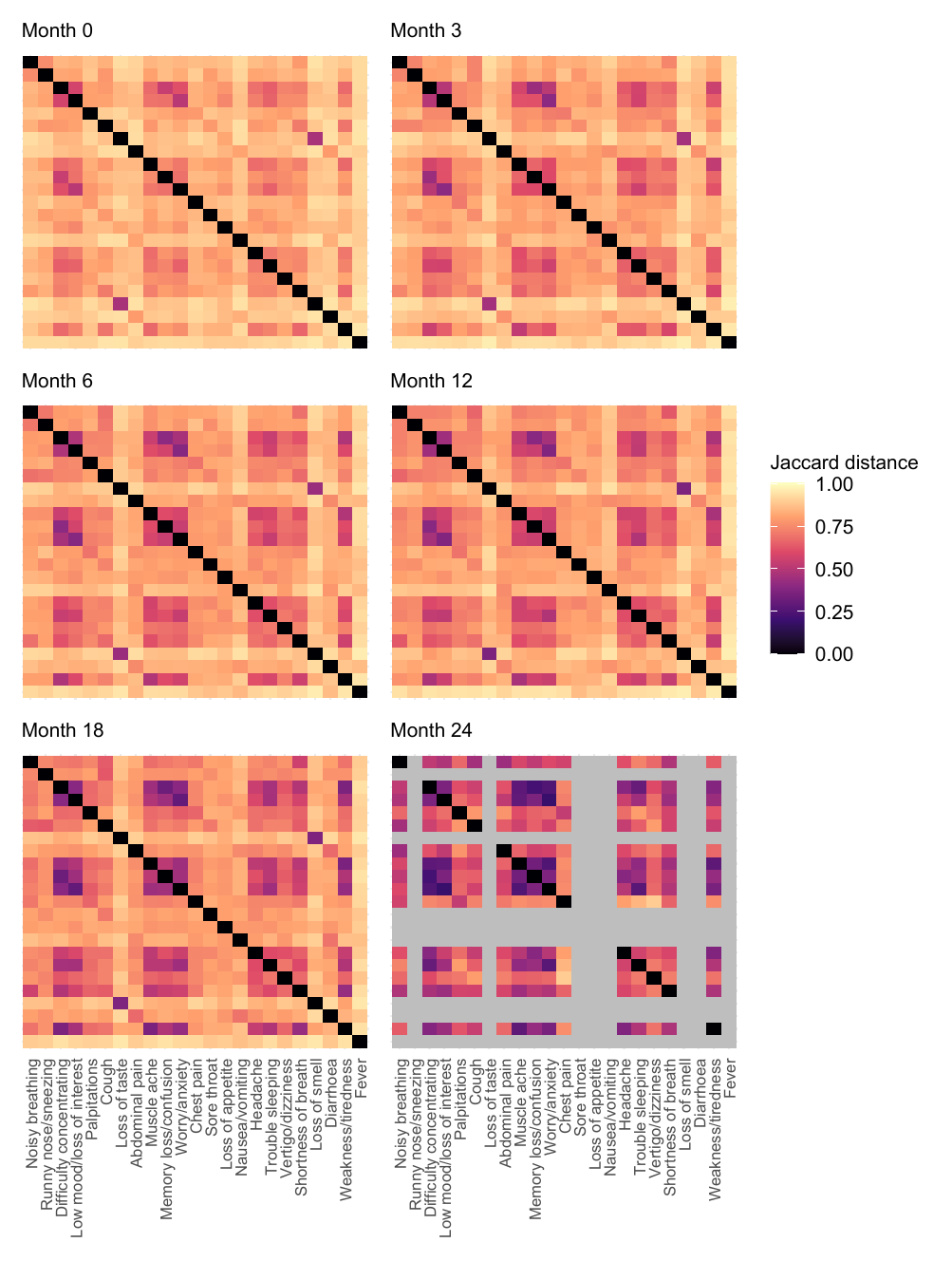}
    \caption{Jaccard distance matrices visualised as heatmaps between symptoms for the second part of the analysis. The included survey responses have been stratified by participants' month-since-onset of long COVID as indicated by the month number at the top of each plot. The following symptoms were removed from the 24 months-since-onset plot due to values suppressed by statistical disclosure control rules: fever, diarrhoea, loss of smell, nausea/vomiting, loss of appetite, sore throat, loss of taste, and runny nose/sneezing.}
    \label{Figure3}
\end{figure*}

We first visualise the Jaccard distances between symptoms using heatmaps. Figures \ref{Figure2} and \ref{Figure3} show the results for the first analysis using every visit with a positive long COVID response and the second analysis which stratifies by time-since-onset of long COVID respectively. Cells in the heatmap are mapped to a black-to-yellow spectrum from symptoms that are always experienced together (with a distance of 0) to symptoms that are never experienced together (with a distance of 1). As expected from the definition of the Jaccard distance in Equation \ref{Equation1}, the diagonal shows perfect self-agreement between symptoms. The relationships between symptoms in the remaining off-diagonal space can be read by tracing up to the diagonal from a symptom cell and then tracing across to the corresponding cell to identify the paired symptom.

Across all of the plots we see that fever is largely experienced quite independently of other symptoms and loss of taste and loss of smell are largely experienced only with each other. From Figure \ref{Figure3} we find higher overall symptom co-occurrence as time-since-onset of long COVID increases.

\subsection{Non-metric multidimensional scaling}
Secondly, we conducted NMDS, an ordination technique used to represent high dimensional data in a low dimensional space using the rank order of distances, on the Jaccard distance matrices of symptoms \cite{Kruskal1964theory,Kruskal1964method}. To facilitate interpretation, we restricted the NMDS ordination to two dimensions resulting in two-dimensional plots where symptoms with high co-occurrence are placed closer together than those with low co-occurrence. 

\begin{figure*}[htb]
    \centering
    \includegraphics[width=0.9\textwidth, keepaspectratio]{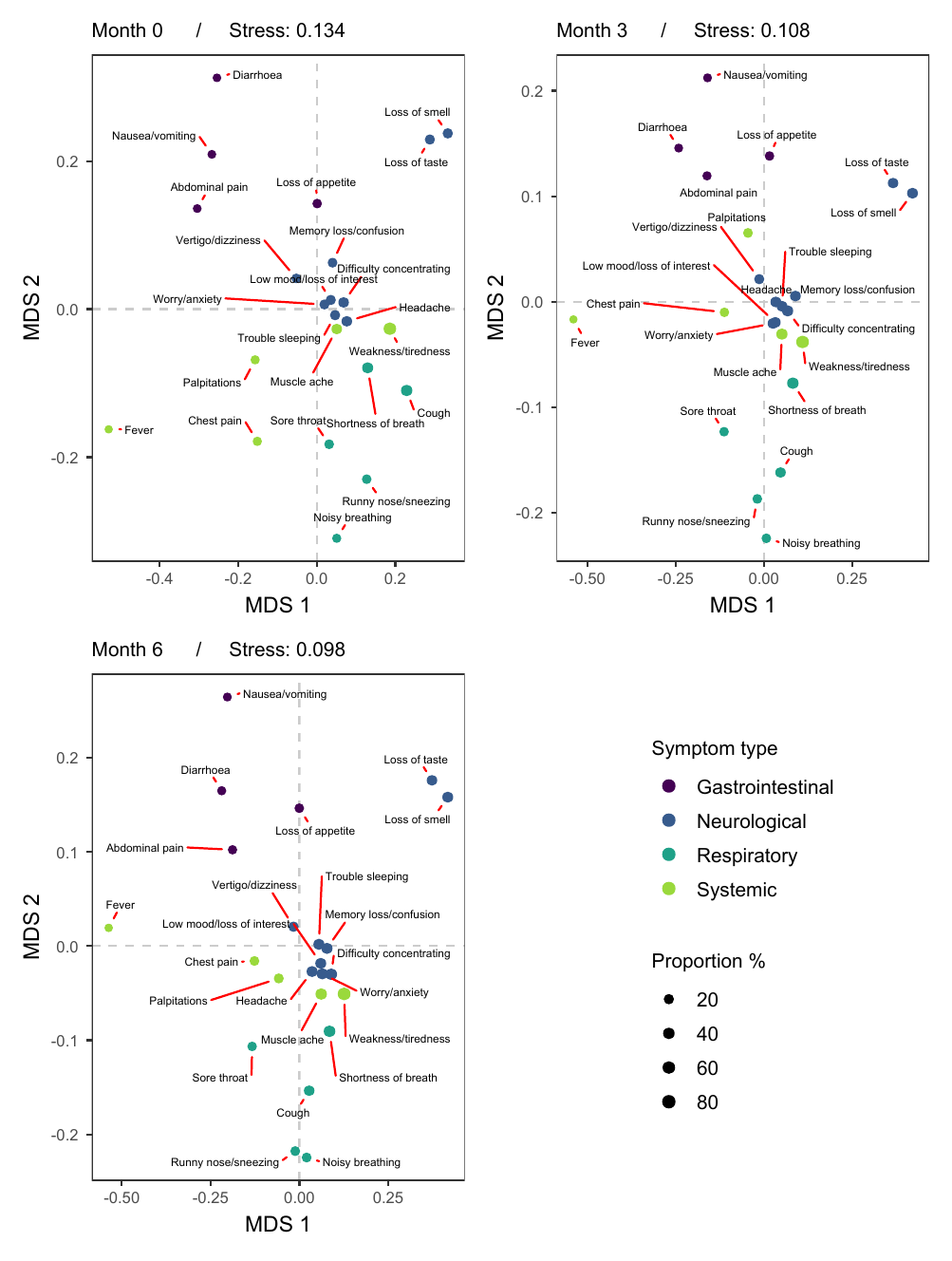}
    \caption{Non-metric multidimensional scaling results for responses from zero to six months-since-onset. The included survey responses have been stratified by participants' month-since-onset of long COVID as indicated by the month number at the top of each plot. Points are colour-coded by symptom type and sized to represent the proportion of participants who experienced said symptom.}
    \label{Figure4}
\end{figure*}

\begin{figure*}[htb]
    \centering
    \includegraphics[width=0.9\textwidth, keepaspectratio]{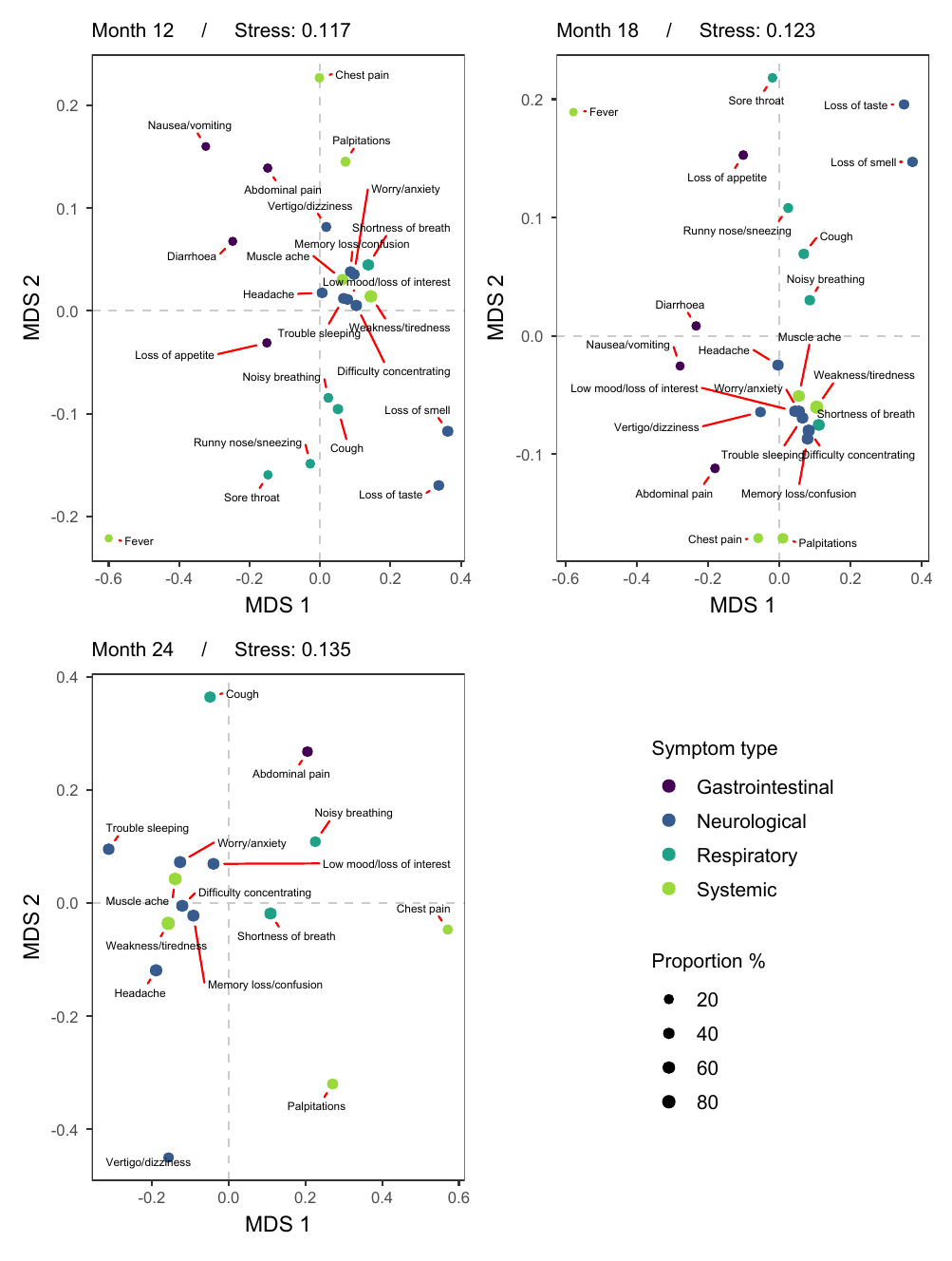}
    \caption{Non-metric multidimensional scaling results for responses from 18 to 24 months-since-onset. The included survey responses have been stratified by participants' month-since-onset of long COVID as indicated by the month number at the top of each plot. Points are colour-coded by symptom type and sized to represent the proportion of participants who experienced said symptom. Some symptoms were removed from the 24 months-since-onset plot due to values suppressed by statistical disclosure control rules.}
    \label{Figure5}
\end{figure*}

The resulting plots can be found in Figures \ref{Figure2}, \ref{Figure4}, and \ref{Figure5} and we note that all plots have acceptable stress values indicating reliable ordination. Symptom points in the plots are colour-coded by symptom type and sized to represent the proportion of participants who experienced said symptom.

Similarly to as in the heatmaps we see that fever is commonly experienced by itself and loss of taste and loss of smell are commonly experienced together too. Across all of the plots we see a central cluster of neurological/systemic symptoms including those relating to altered mental state, sleep, muscle ache, and weakness/tiredness. Across the time-since-onset plots in Figures \ref{Figure4} and \ref{Figure5} we find separate clusters of gastrointestinal symptoms and respiratory symptoms on either side of the central cluster. At month 18 this separation of symptoms by organ system becomes less pronounced but the central cluster persists. 

From the proportions in all of the NMDS plots, we immediately see that fever is the least commonly experienced symptom. We find that weakness/tiredness is the most commonly experienced symptom over each month-since-onset plot in Figures \ref{Figure4} and \ref{Figure5}, but is not the most commonly reported when looking at all of the responses in Figure \ref{Figure2}. Similarly, shortness of breath becomes more commonly experienced as time-since-onset increases, but is one of the least reported symptoms when looking at all of the responses. The neurological/systemic symptoms appear to be more common than the gastrointestinal symptoms in the time-since-onset plots, however, loss of appetite is the most experienced symptom when looking at all of the responses. Finally, as in the heatmaps, the data shows that the proportions of participants experiencing each symptom increases as time-since-onset increases, again indicating higher overall symptom co-occurrence in later long COVID.

\subsection{Agglomerative hierarchical clustering}
\begin{figure*}[htb]
    \centering
    \includegraphics[width=0.9\textwidth, keepaspectratio]{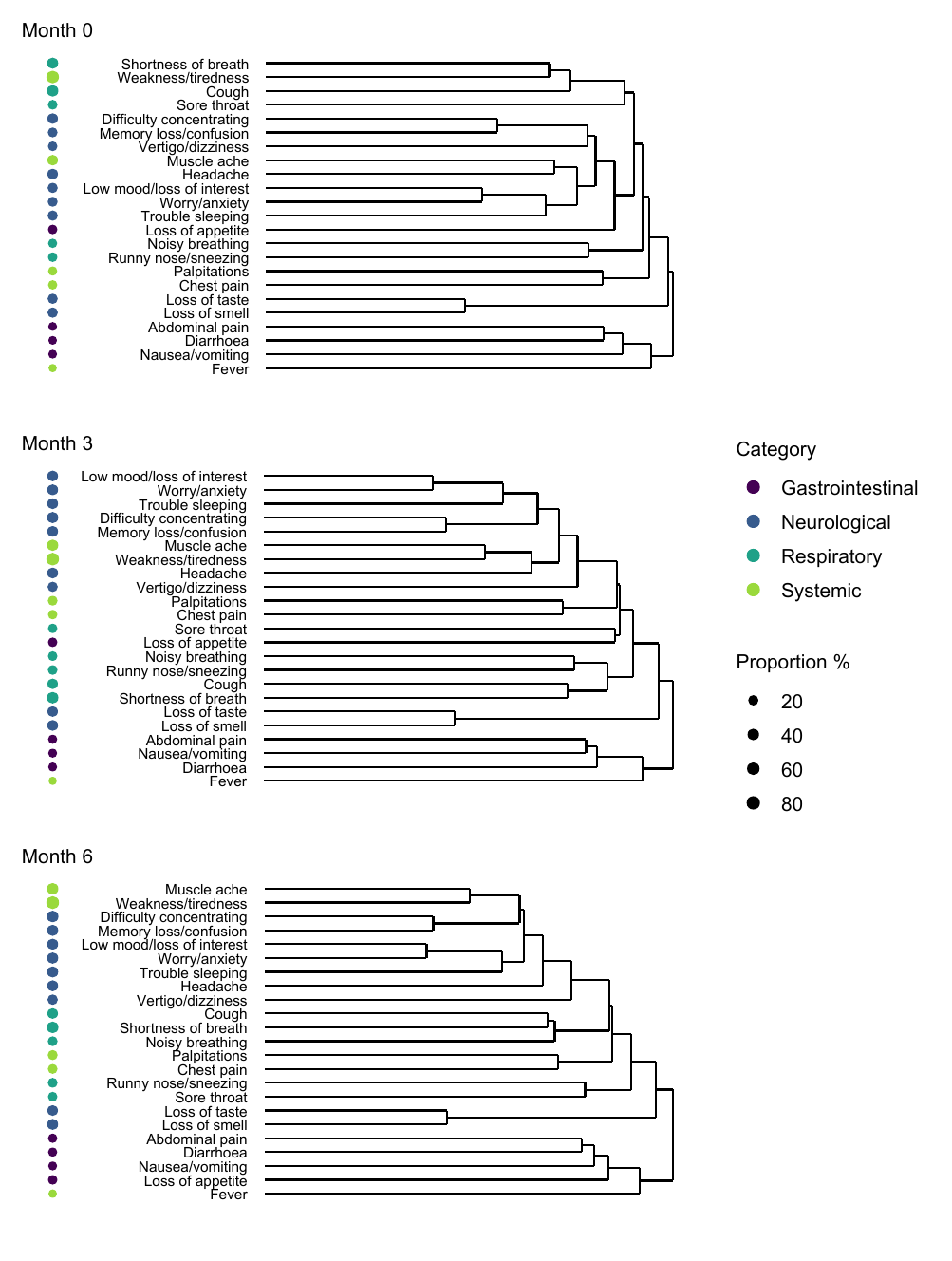}
    \caption{Dendrograms visualising agglomerative hierarchical clustering results for responses from zero to six months-since-onset. The included survey responses have been stratified by participants' month-since-onset of long COVID as indicated by the month number at the top of each plot. Points are colour-coded by symptom type and sized to represent the proportion of participants who experienced said symptom. Agglomerative hierarchical clustering was performed using complete linkage.}
    \label{Figure6}
\end{figure*}

\begin{figure*}[htb]
    \centering
    \includegraphics[width=0.88\textwidth, keepaspectratio]{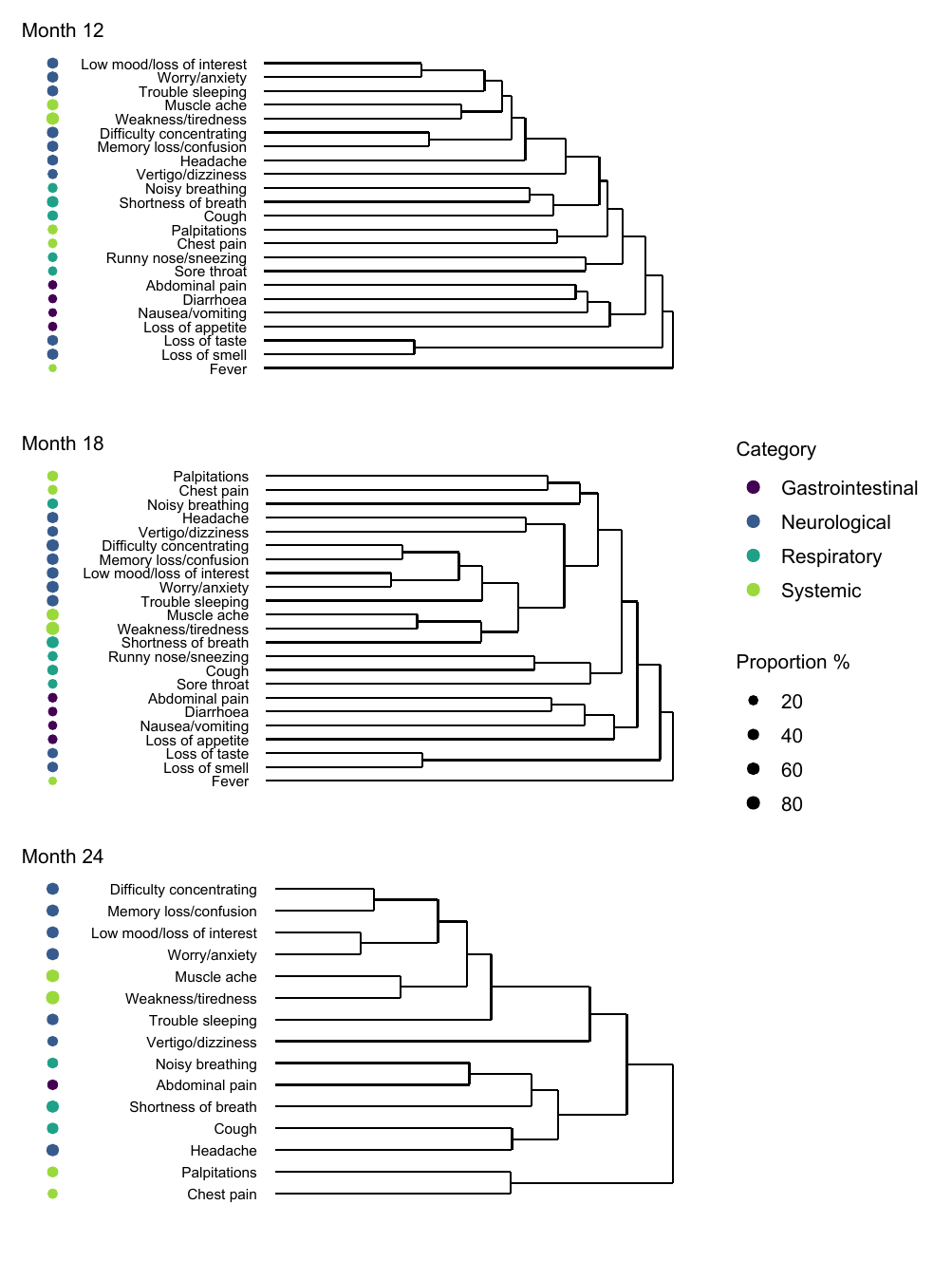}
    \caption{Dendrograms visualising agglomerative hierarchical clustering results for responses from 18 to 24 months-since-onset. The included survey responses have been stratified by participants' month-since-onset of long COVID as indicated by the month number at the top of each plot. Points are colour-coded by symptom type and sized to represent the proportion of participants who experienced said symptom. Some symptoms were removed from the 24 months-since-onset plot due to values suppressed by statistical disclosure control rules. Agglomerative hierarchical clustering was performed using complete linkage.}
    \label{Figure7}
\end{figure*}

Lastly, we performed agglomerative hierarchical clustering on the Jaccard distances matrices using complete linkage \cite{James2021}. Figures \ref{Figure2}, \ref{Figure6}, and \ref{Figure7} show the dendrograms resulting from this clustering technique and are read from right to left with symptoms that are joined further to the left co-occurring more frequently according to the Jaccard distance. Symptoms are labelled with the same points from the NMDS plots that are colour-coded by symptom type and sized to represent the proportion of participants who experienced said symptom.

Across most of the dendrograms we see the leftest joining symptoms are the pairs of loss of smell and loss of taste, difficulty concentrating and memory loss/confusion, and low mood/loss in interest and worry/anxiety. As time increases across Figures \ref{Figure6} and \ref{Figure7} muscle ache and weakness/tiredness are joined further to the left. With proportions of 69\% and 80\% respectively at 24 months-since-onset this indicates these become core symptoms in those who experience long COVID for a prolonged time. Furthermore, we see evidence of chaining particularly between months 3 and 12 since-onset. Chaining is where the clustering leads to long, thin clusters and symptoms are joined to a main cluster one by one rather than being allocated into distinct, balanced clusters in the dendrograms. At month 18 in particular in Figure \ref{Figure7}, and in Figure \ref{Figure2} of all of the long COVID responses, we see a distinct cluster of neurological symptoms with some systemic symptoms.

\section{Discussion}

Our analysis in the CIS investigated long COVID symptom co-occurrence. We presented three methods to visualise this: heatmaps, NMDS, and agglomerative hierarchical clustering with complete linkage. Each method used Jaccard symptom by symptom distance matrices derived from binary survey responses of symptom presence. We split our analysis into two parts, first looking at all long COVID survey responses and then looking at responses stratified by time-since-onset of long COVID up to \(24\) months. 

All three visualisation methods, across both analyses, illustrated that fever rarely occurs with other symptoms and is the least frequently experienced symptom. Fever is more frequently experienced during an active infection \cite{Evans2015}; since long COVID is a post-viral condition \cite{NICE_LC, WHO2021}, this lower frequency of fever makes sense. We also found that loss of taste and smell co-occur in a distinct, independent cluster. This suggests that some people may experience prolonged loss of taste and smell post acute COVID-19 infection without suffering from any other long COVID symptoms. 

The data further revealed clear changes in symptom presentation as time-since onset of long COVID increases. The heatmaps (Figure \ref{Figure3}) demonstrate higher overall symptom co-occurrence through time, while the chaining we see in the agglomerative hierarchical clustering dendrograms (Figures \ref{Figure6} and \ref{Figure7}) in earlier long COVID suggests a lack of distinct subgroups of symptoms. Together these suggest that early long COVID can be characterised by a highly heterogeneous presentation of symptoms, while later long COVID is associated with multi-organ system symptoms and increased symptom burden. This increased symptom burden can also be seen in Table \ref{Table2}. Furthermore, from the agglomerative hierarchical clustering dendrograms we see that muscle ache and weakness/tiredness co-occur increasingly more frequently over time-since-onset (as well as being highly prevalent symptoms), suggesting these are core symptoms in those who experience long COVID for a prolonged time. This is further evidenced in the NMDS plots where we observe symptoms of muscle ache and weakness/tiredness, alongside those relating to altered mental state and sleep, become an increasingly tighter cluster as time-since-onset increases. We note that the NMDS plot from 24 months-since-onset has a different scale to the other plots due to the ordination being performed only on symptoms that were not suppressed by statistical disclosure control rules. Something that remains consistent throughout the increasing time-since-onset NMDS plots is that shortness of breath remains close to the main cluster of symptoms, with a consistently high proportion of participants experiencing it, suggesting this symptom is also a core symptom of long COVID.

The final key insight from our analysis is the clear separation between neurological/systemic, gastrointestinal, and respiratory symptoms. This divergence is observable across the whole analysis but is most notable in the early stages of long COVID, suggesting that these symptom groups could represent symptom phenotypes of early long COVID. Also notably, the potential gastrointestinal and respiratory symptom phenotypes were reported less frequently than the neurological/systemic symptom phenotype.

Our findings are consistent with the broader literature. The data revealed symptom co-occurrence clusters consistent with the potential patient phenotypes and organ system clusters as found in the systematic review by \citet{Wang2026}. Similarly, we found the same emergence of multi-system symptoms in prolonged long COVID as highlighted in the systematic review by \citet{Hu2025}, as well as potential neurological, respiratory, and gastrointestinal phenotypes, and the emergence of fatigue as a core symptom. Conversely, our findings also diverge as we do not see an obvious cluster of circulatory symptoms.

Beyond the symptom co-occurrence patterns and although not detailed in the results, Table \ref{Table2} warrants closer attention. Specifically that relating to the number of COVID-19 vaccinations received by those experiencing long COVID 24 months-since-onset. We find that none of the participants who were experiencing long COVID 24 months-since-onset received more than one COVID-19 vaccination. We do note that, since the survey ended in March 2023, these people's long COVID experiences would have begun in March 2021, however, vaccine rollout in the UK begun on 8 December 2020 for those in the highest priority groups and was not available to all adults until 18 June 2021. This suggests that these people may not have been able to receive a vaccination before contracting SARS-COV-2, however, they did not then receive more than one vaccination after developing long COVID. in fact we see that the majority of participants at each stratification by time-since-onset received no vaccinations. Research in the CIS has shown that vaccination can reduce the risk of developing long COVID \cite{Ayoubkhani2022_OFID} and can reduce symptoms in those who have already developed long COVID \cite{Ayoubkhani2022_BMJ}. Consequently, this highlights an area for further research.

A notable limitation of our analysis in this paper is its dependence on self-reporting. As symptoms such as fatigue and headaches are historically common somatic complaints among the general population \cite{Stadje2016, Husy2025}, reliance on self-reporting risks false-positive identifications of these symptoms being a result of long COVID. While the long COVID question in the CIS was designed to mitigate this by instructing participants to report only symptoms not attributable to anything else, misclassification cannot be entirely ruled out. Furthermore, self-reporting symptoms could have introduced recency bias, leading to under-reporting, since the survey was administered in general monthly for having had any of the symptoms. Additionally, there is the possibility of both under-reporting and over-reporting due to subjective thresholds over what counts as the presence of a symptom. These biases could affect the reliability of our data. Participation bias is also a limitation in this analysis. While the survey was designed to be nationally representative, actual enrolment was skewed with significantly lower response rates among younger age groups and within certain ethnic minority communities \cite{ONS_CIS_QMI}. Furthermore, those who voluntarily enrol in a long-term government health tracking study may behave differently from the general public \cite{ONS_CIS_QMI}. Lastly, the CIS sampled entire households, which could have skewed the data towards specific family behaviours rather than purely independent individuals. This participation bias risks compromising the data representativeness to the general UK population. For our methodology, a limitation does exist in using the Jaccard distance. The Jaccard distance metric is not prevalence-adjusted; this means that highly prevalent symptoms can exhibit lower distances due to their high baseline rates of occurrence rather than a true clinical association, which should be considered during interpretation. Finally, the second analysis is complementary to the first analysis, which included all responses with a positive long COVID response. In the first analysis, multiple responses from a single participant may be overrepresented compared to those who contributed only a single response. The second analysis thereby increases confidence that our findings are robust.

While these participation biases must be acknowledged, a major strength of this analysis remains the use of the CIS. Its selection of households from a representative population, large sample size, and prospective community-based cohort design provide methodological advantages and reduce biases common in retrospective and convenience-sampled study designs \cite{Browner2023}. Furthermore, the ONS maintained telephone alternatives for participants unable to access the survey online during the transition to remote data collection in July 2022 \cite{Oxford_protocols}, which helped to mitigate digital exclusion and preserved survey representativeness. In addition, the self-reporting approach of each participant's long COVID experience allows us to fully capture the heterogeneous experiences of long COVID that might not be otherwise. By requiring a clinical confirmation of long COVID, or a confirmed acute COVID-19 infection, many people would have been excluded. This would have been especially evident for survey responses early in the pandemic, where COVID-19 testing in the UK was limited \cite{Oxford_protocols}. The continued lack of an unified definition of long COVID would also affect this \cite{Gomez-Bravo2025}. Analysis with the self-reported long COVID experiences allows us to capture participants facing barriers to healthcare access and those who did not seek a formal diagnosis. Crucially this enables us to see all experiences of long COVID from mild to severe and to represent the heterogeneity of long COVID symptoms. Moreover, the use of a binary (yes/no) metric to symptoms allows us to capture whether a wide range of symptoms are present simply. Lastly, we think our methods provide a good balance between sophistication and interpretability.

Due to the variability of long COVID, defining severity of the condition can be challenging. Since our work shows that in those who experience prolonged long COVID, that symptom co-occurrence increases, we would suggest using the number of symptoms to contribute to the definition of the severity of later long COVID. This aligns with results from a statistical bulletin published by the ONS from the CIS showing direct correlation between total symptom burden and functional daily impairment \cite{ONS_LC_July23}.

\section{Conclusion}

We conclude that research into long COVID and other post viral conditions needs to remain a public health priority given the significant impact it has on the healthcare system, the economy, and most importantly the individuals living with it. Pandemic preparedness frameworks also need to reflect this. We have demonstrated that the symptom experience of long COVID evolves from early stages to late stages with higher symptom burden and multi-systemic presentation. This indicates that the severity of later long COVID could be at least partially characterised by the number of symptoms being experienced. We find three possible symptom phenotypes of early long COVID from neurological/systemic, respiratory, and gastrointestinal symptom groups, which is consistent with the broader literature. Our work points to future research directions in vaccination/immunology and symptom cluster trajectories.

\section*{List of abbreviations}
\begin{itemize}
    \item COVID-19: Coronavirus Disease 2019
    \item SARS-CoV-2: Severe acute respiratory syndrome coronavirus 2
    \item OECD: Organisation for Economic Co-operation and Development
    \item CIS: COVID-19 Infection Survey
    \item ONS: Office for National Statistics
    \item UK: United Kingdom
    \item NMDS: Non-metric multidimensional scaling
\end{itemize}

\section*{Declarations}

\subsection*{Ethics approval and consent to participate}
Ethical approval for the CIS was received from the South Central Berkshire B Research Ethics Committee (20/SC/0195). Written informed consent was obtained for all participants aged 16 years or older. Written informed consent was obtained from parents or carers for participants aged 2-15 years, with written assent also provided from those aged 10-15 years.

\subsection*{Consent for publication}
Captured in the CIS study protocol.

\subsection*{Availability of data and materials}
The data that support the findings of this study are available from the Office for National Statistics but restrictions apply to the availability of these data. De-identified study data are available to accredited researchers in the ONS Secure Research Service under part 5, chapter 5 of the Digital Economy Act 2017. Data that received SRS output clearance and custom code written to support the work in this article are available via \url{https://doi.org/10.5281/zenodo.22251745} \cite{GitAherne}.

\begin{itemize}
    \item \textbf{Project name:} LC\_SymptomClustersPublic
    \item \textbf{Project home page:} Accessible via \url{https://github.com/jasmineaherne/LC_SymptomClustersPublic}
    \item \textbf{Archived version:} \url{https://doi.org/10.5281/zenodo.22251745}
    \item \textbf{Operating system(s):} Platform independent
    \item \textbf{Programming language:} R
    \item \textbf{Other requirements:} R version 4.3.0 or higher
    \item \textbf{License:} MIT Licence
    \item \textbf{Any restrictions to use by non-academics:} None
\end{itemize}

\subsection*{Competing interests}
The authors declare that they have no competing interests.

\subsection*{Funding}
JA is supported by an EPSRC Doctoral Training Award. TH is supported by the Wellcome Trust (Grant number 227438/Z/23/Z), the Medical Research Council (Grant number UKRI483), and UKRI (Grant number MR/Z505328/1).

\subsection*{Authors' contributions}
Conceptualisation: all authors; Methodology: all authors; Formal analysis: JA; Writing - original draft preparation: JA; Writing - review and editing: all authors; Funding acquisition: TH.

\subsection*{Acknowledgements}
This work was undertaken in the Office for National Statistics Secure Research Service using data from ONS and other owners and does not imply the endorsement of the ONS or other data owners.


\end{document}